\documentstyle[12pt]{article}
\newcommand{\eut}{
\begin{picture}(3,3)(-2,-2)
\put(-2,-15){$\tilde{}$}
\put(-5,-2){$\eta$}
\end{picture}}

\newcommand{\nut}{
\begin{picture}(11,5)(0,0)
\put(5,-10){$\tilde{}$}
\put(0,0){\it N}
\end{picture}}

\newcommand{\mut}{
\begin{picture}(12,5)(0,0)
\put(6,-10){$\tilde{}$}
\put(0,0){\it M}
\end{picture}}

\newcommand{\qu}{
\begin{picture}(8,5)(0,0)
\put(4,0){$\tilde{}$}
\put(4,2){$\tilde{}$}
\put(0,0){\it q}
\end{picture}}

\begin{document}
\baselineskip=22pt plus 0.2pt minus 0.2pt
\lineskip=22pt plus 0.2pt minus 0.2pt
\begin{center}
 \Large
The Phase Space of 2+1 Dimensional\\
Gravity in the Ashtekar Formulation\\

\vspace*{0.35in}

\large

J.\ Fernando\ Barbero\ G. $^{1,2}$ and
Madhavan Varadarajan $^{1}$
\vspace*{0.25in}

\normalsize

$^{1}$Department of Physics, Syracuse University, \\
Syracuse, New York 13244-1130, U.S.A.
 \\
$^{2}$Instituto de Matematicas y F\'{\i}sica Fundamental, \\
C.S.I.C.\\
Serrano 119--123, 28006 Madrid, Spain
 \\

\vspace{.5in}
June 14, 1993\\
\vspace{.5in}
ABSTRACT

\end{center}

The Ashtekar formulation of 2+1 gravity differs from the geometrodynamical and
Witten descriptions when the 2-metric is degenerate. We study the phase space
of 2+1 gravity in the Ashtekar formulation to understand these degenerate
solutions to the field equations. In the process we find two new systems of
first class constraints which describe part of the degenerate sectors of the
Ashtekar formulation. One of them also generalizes the Witten constraints.
Finally we argue that the Ashtekar formulation has an arbitrarily large number
of degrees of freedom in contrast to the usual descriptions.
\pagebreak

\setcounter{page}{1}

\section{Introduction} In the  last few years, a great deal of effort has been
devoted to the problem of understanding the canonical quantization of gravity
following the path opened up by Ashtekar with the introduction of a new set of
variables to describe the gravitational field\cite{1}\cite{2}. Although the
program has given new insights
and has addressed (and solved) important problems, there still
remains a great deal of work to be done in order to arrive at a consistent
quantum theory of gravity.

As a way of gaining some insight into how
the program of quantization of 3+1 gravity can be completed, the study of toy
models which retain some of the characteristics of the full theory is of much
help. One of these models is 2+1 gravity. Since it is generally covariant, we
can learn how to deal with the problems related to diffeomorphism invariance
faced by the quantization program in 3+1 gravity. One of the motivations of
this
work is to study 2+1 gravity as a toy model for the Ashtekar formulation of
3+1 gravity.

At this point we would like to emphasize that the constraints of 2+1 vacuum
gravity can be written in several ways. Two of these, namely, the description
in terms of geometrodynamic variables \cite{100} and Witten's reformulation of
the theory \it $\grave{a}$ la \rm Chern--Simons \cite{4} \cite{3}
(much in the spirit of Ashtekar's proposal of using a connection as the
fundamental variable to describe gravity instead of the metric) have been
studied in great detail in the literature. In this paper we will use another
set of constraints for 2+1 gravity that closely mimics the more complicated
ones of 3+1 gravity written in the Ashtekar variables. The reason for
this is that, as already stated, we would like to learn something about the
Ashtekar program in 3+1 gravity.

The fact that 2+1 gravity can be described by
constraints similar in form to the Ashtekar ones in 3+1 dimensions
 was first pointed out by
Bengtsson \cite{5} who proved that when the frame field is non--degenerate
both systems of constraints are equivalent. However, the degenerate frame
fields play a very different role in the Witten and Ashtekar
formulations of the theory. Though the situation in the first case is clearly
understood the same is not true for the second. In this paper we want to
clarify the role of the degenerate frame fields in Ashtekar's description
of 2+1 gravity in order to
understand their role in the more relevant 3+1 case. As has been shown in
\cite{6} it is possible to have non-singular negative energy solutions to the
Ashtekar constraints in 3+1 dimensions if degenerate metrics are permitted, so
it is important to know what their role is in the quantum description.

The second motivation for this work is tied to the realization that 3+1 gravity
with one hypersurface orthogonal spacelike Killing vector field is equivalent
to 2+1 gravity coupled to a massless scalar field. Hence the study of 2+1
gravity with a certain type of matter is actualy a study of a midisuperspace of
Einstein's theory. One cannot couple local matter fields to Witten's
constraints of 2+1 gravity. However one can couple matter polynomially if one
works within Ashtekar's framework.
Of course, before studying matter couplings, it is necessary to
understand the theory without matter.

In what follows we will concentrate on the study of the constraints of
2+1 gravity  in the Ashtekar formulation. The "spacetime" manifold is
diffeomorphic to $\Sigma \times R$, where $\Sigma$ is a compact, orientable
2-manifold without boundary.
The lay out of the paper is as follows.
Section 2 is devoted to the study of two new systems of constraints that
describe sectors of the Ashtekar description of 2+1 gravity. They can be found
by concentrating on the study of null curvatures. The first of these two
systems
is a generalization of the Witten constraints but contains only part of
the degenerate sectors of Ashtekar's theory whereas the second describes the
degenerate solutions contained in Ashtekar's description of 2+1 gravity.
The Poisson algebra is simpler in the new constraints (though by no means
trivial!).

In the third section we study some features of the dynamics of the degenerate
solutions; we will see that in the Ashtekar case, the theory may have an
infinite number of degrees of freedom. This is in sharp contrast with what
happens in the Witten formulation for which the number of degrees of freedom
is finite (and depends only on the topology of the 2- manifold). Null
curvatures play an special role here too.

In section 4 we give an explicit example of a solution to the
Ashtekar constraints which clearly shows that the theory has an arbitrarily
large
number of degrees of freedom. We end the paper with  conclusions and some
open questions.

\section{The New Constraints}

We start by introducing our conventions and notation. The configuration
variable for 2+1 gravity is a real $SO(2,1)$ valued connection $A_{a}^{I}$
with conjugate momentum $\tilde{E}^{a}_{I}$ (the frame fields or triads).
In the following a, b, c, etc.
(running from 1 to 2) will represent tangent space indices; internal indices
will be denoted by I, J, K, etc (running from 1 to 3). They are raised and
lowered with the (internal) Minkowski metric $\eta_{IJ}$ with signature (--, +,
+). The Levi-Civita
tensor density and its inverse will be represented by $\tilde{\eta}^{ab}$ and
$\eut_{ab}$ respectively. The usual convention of representing the
density weight of an object with tildes above or below the fields (positive
and negative density weights respectively) will be used throughout the paper.
The covariant derivatives are given by
${\cal D}_{a}\alpha_{I}=\partial_{a}\alpha_{I}+\epsilon_{IJ}^{\;\;\;\;K}
A_{a}^{J}\alpha_{K}$, the curvature is $F_{abI}=2\partial_{[a}A_{b]I}+
\epsilon_{I}^{\;\;JK}A_{aJ}A_{bK}$, where $\epsilon^{IJK}$ is the internal
Levi-Civita tensor. Finally the Poisson brackets between the
connection and frame fields are $\{A_{a}^{I}(x), \tilde{E}^{b}_{J}(y)\}=
\delta^{2}(x,y)\delta^{\;\;b}_{a}\delta^{\;\;I}_{J}$

\noindent Witten  has shown in \cite{4} that 2+1 gravity is described by the
following system of constraints:

\begin{eqnarray}
& & {\cal D}_{a} \tilde{E}^{a}_{I}=0\nonumber\\
& & F_{ab}^{I}=0\label{1}
\end{eqnarray}

Another set of constraints for 2+1 gravity are analogous to the Ashtekar
ones in 3+1 gravity \cite{5}:

\begin{eqnarray}
& & {\cal D}_{a} \tilde{E}^{a}_{I}=0\nonumber\\
& & \tilde{E}^{b}_{I} F_{ab}^{I}=0\label{2}\\
& & \epsilon^{IJK}\tilde{E}^{a}_{I}\tilde{E}^{b}_{J}F_{abK}=0\nonumber
\end{eqnarray}

\noindent They are called the Gauss, vector and scalar or hamiltonian
constraint respectively. Both (\ref{1}) and (\ref{2}) are first class systems
and when the triads are non--degenerate they are equivalent \cite{5}. The fact
that we have six constraints and six configuration variables per point
indicates via naive counting that there may be no local degrees of freedom.
Following Bengtsson \cite{5} we
may write $F_{ab}^{I} = \tilde{v}^{I}\eut_{ab}$ where
$\tilde{v}^{I}$ is an internal vector density of weight +1 and then
(\ref{2}) become:

\begin{eqnarray}
& & {\cal D}_{a} \tilde{E}^{a}_{I}=0\nonumber\\
& & \tilde{E}^{a}_{I}\tilde{v}^{I}\eta_{ab}=0\label{2b}\\
& & \epsilon^{IJK}\;\eut_{ab}\tilde{E}^{a}_{I}\tilde{E}^{b}_{J}
\tilde{v}_{K}=0\nonumber
\end{eqnarray}

These equations admit the following interpretation: Suppose, we take
solutions to the Gauss constraint; then the vector constraint tells us that the
triads must be orthogonal to $\tilde{v}^{I}$ in the internal index. The
hamiltonian constraint is simply the statement that $\tilde{E}_{I}^{1}$,
$\tilde{E}_{I}^{2}$ and $\tilde{v}_{K}$ must be linearly dependent. We must
discuss now three different possibilities:

1) If $\tilde{v}_{K}$ is timelike, the vector and scalar constraints are solved
by $\tilde{E}_{I}^{1}$, $\tilde{E}_{I}^{2}$ spacelike, orthogonal to
$\tilde{v}_{K}$ and colinear.

2) If $\tilde{v}_{K}$ is spacelike the vector and scalar constraints are solved
by finding $\tilde{E}_{I}^{1}$, $\tilde{E}_{I}^{2}$ orthogonal to
$\tilde{v}_{K}$ and colinear; in this case the frame fields can be spacelike,
timelike or null.

3) If $\tilde{v}_{K}$ is null then the vectors orthogonal to it are contained
in the null plane orthogonal to
$\tilde{v}_{K}$. Now $\tilde{v}_{K}$ and the triads are always linearly
dependent so that, once we have solved the vector constraint, the scalar
constraint is identically satisfied. This suggests that we can substitute the
scalar constraint by the condition that $\tilde{v}_{K}$ is null and get the
following set of constraints that capture part of the information in the
degenerate sector of Ashtekar's theory:

\begin{eqnarray}
& & {\cal D}_{a} \tilde{E}^{a}_{I}=0\nonumber\\
& & \tilde{E}^{b}_{I} F_{ab}^{I}=0\label{2c}\\
& & \tilde{\eta}^{ab}\tilde{\eta}^{cd}F_{ab}^{I}F_{cdI}=0\nonumber
\end{eqnarray}


We must check, of course, that these constraints are
first class but this is straightforward because the new
hamiltonian constraint is independent of momenta.

In the previous discussion we have seen  that any solution to the set of
constraints (\ref{2b}) is also a
solution to the Ashtekar constraints for 2+1 gravity; obviously a solution to
Witten's constraints is also a solution to (\ref{2b}). We have,then,
 a new set of
constraint equations that are more general than the Witten ones (they
contain degenerate solutions in a non--trivial way) but capture only part of
the degenerate sectors of the Ashtekar formulation of 2+1 gravity. Furthermore
the constraints (\ref{2c}) retain some of the nice features of the previous
formulations; for example they are polynomial in the canonical variables.

In the formulation given by (\ref{2}) one can have a non flat connection only
when the triads are degenerate. From our analysis it is possible to check that
if $\tilde{v}_{K}$ is spacelike or timelike the constraints can be solved
only if the triads $\tilde{E}_{I}^{1}$, $\tilde{E}_{I}^{2}$ are proportional
to each other so that the degeneracy of the metric is guaranteed. In the null
case, on the other hand, the triads need not be colinear to satisfy the
constraints, however the metric is still degenerate. In order to see this, let
$s_{K}$ be a unit non null internal vector orthogonal to  $\tilde{v}_{K}$
and let $v_K$ be proportional to
$\tilde{v}_{K}$. Then if $e_{aK}\;\;\;a=1,2$ are vectors orthogonal to
$\tilde{v}_{K}$ they can be written as $e_{aK}=\mu_{a}v_{K}+ \lambda_{a}
s_{K}$ for a suitable choice of $s_{K}$.
Then $e_{aK}e_{b}^{\;\;K}$ has the form:

\[ \left[ \begin{array}{cc}
\lambda_{1}^{2} & \lambda_{1}\lambda_{2}\\ \lambda_{1}\lambda_{2} &
\lambda_{2}^{2}
\label{3}
\end{array} \right] \]

\noindent We see then why $\qu^{ab}=\tilde{E}^a_I\tilde{E}^{bI}$ is degenerate
if $\tilde{v}_{K}$ is null even though the triads are not colinear.

In order to study the gauge transformations generated by the constraints
(\ref{2c}) one introduces the following constraint functionals:

\begin{eqnarray}
& & G(\Lambda^{I})=\int_{\Sigma}d^{2}x\; \Lambda^{I}{\cal D}_{a}
\tilde{E}^{a}_{I}\label{4}\\
& & D(N^{a})=\int_{\Sigma}d^{2}x N^{a}[F_{ab}^{I}\tilde{E}_{I}^{b}-A_{a}^{I}
{\calD}_{b}\tilde{E}^{b}_{I}]\label{5}\\
& & S(\nut)=\frac{1}{2}\int_{\Sigma}d^{2}x \nut
\tilde{\eta}^{ab}F_{ab}^{I}\tilde{\eta}^{cd}F_{cdI}\label{6}
\end{eqnarray}

\noindent We then find:

\begin{eqnarray}
& & \left\{G(\Lambda^{I}),A_{a}^{I}\right\}=-{\cal D}_{a} \Lambda^{I}
\label{7}\\
& & \left\{G(\Lambda^{I}),\tilde{E}^{a}_{I}\right\}=-\epsilon_{I\;\;K}^{\;\;J}
\tilde{E}^{a}_{J}\Lambda^{K}\label{8}\\
& & \left\{D(N^{a}),A_{a}^{I}\right\}={\cal L}_{\vec{N}}A_{a}^{I}
\label{9}\\
& & \left\{D(N^{a}),\tilde{E}^{a}_{I}\right\}
={\cal L}_{\vec{N}}\tilde{E}^{a}_{I}
\label{10}\\
& & \left\{ S(\nut),A^{I}_{a}\right\}=0
\label{11}\\
& & \left\{S(\nut),\tilde{E}_{I}^{a}\right\}=\tilde{\eta}^{ab}{\cal D}_{b}(\nut
\tilde{\eta}^{cd}F_{cdI})
\label{12}
\end{eqnarray}

\noindent Where ${\cal L}_{\vec{N}}$ denotes the Lie derivative along
the vector field $\vec{N}$. The infinitesimal gauge variations generated by
(\ref{4},\ref{5},\ref{6}) are, respectively:

\begin{eqnarray}
& & \left\{ \begin{array}{l}A_{a}^{I}\rightarrow A_{a}^{I}-\delta
{\cal D}_{a}\Lambda^{I}\\
\tilde{E}_{I}^{a}\rightarrow\tilde{E}_{I}^{a}-\delta
\epsilon_{I\;\;K}^{\;\;J}\tilde{E}_{J}^{a}\Lambda^{K}\end{array}\right.
\label{13}\\
& & \left\{ \begin{array}{l}A_{a}^{I}\rightarrow A_{a}^{I}+\delta
{\cal L}_{\vec{N}}A_{a}^{I}\\
\tilde{E}_{I}^{a}\rightarrow\tilde{E}_{I}^{a}+\delta
{\cal L}_{\vec{N}}\tilde{E}^{a}_{I}\end{array}\right.\label{14}\\
& & \left\{ \begin{array}{l}A_{a}^{I}\rightarrow A_{a}^{I}\\
\tilde{E}_{I}^{a}\rightarrow\tilde{E}_{I}^{a}-\delta
\tilde{\eta}^{ab}{\cal D}_{b}(\nut
\tilde{\eta}^{cd}F_{cdI})\end{array}\right.\label{15}
\end{eqnarray}

\noindent Where $\delta$ is an infinitesimal parameter. Finally the Poisson
brackets of the constraints are:

\begin{eqnarray}
& & \{G(\Lambda_{1}),G(\Lambda_{2})\}=G([\Lambda_{1},\Lambda_{2}])\nonumber\\
& & \{G(\Lambda),D(\vec{N})\}=-G({\cal L}_{\vec{N}}\Lambda)\nonumber\\
& & \{G(\Lambda),S(\nut)\}=0\nonumber\\
& & \{D(\vec{N}),D(\vec{M})\}=D([\vec{N},\vec{M}])\label{133}\\
& & \{D(\vec{N}),S(\mut)\}=S({\cal L}_{\vec{N}}\mut)\nonumber\\
& & \{S(\nut),S(\mut)\}=0\nonumber
\end{eqnarray}

\noindent Where $[\Lambda_{1}, \Lambda_{2}]^{I}=\epsilon^{I}_{\;\;JK}
\Lambda_{1}^{J}\Lambda_{2}^{K}$
and the commutator of vector fields is defined as usual. As we can see, at
variance with what happens with the Ashtekar constraints, we have a Lie algebra
of Poisson brackets because we  always have structure constants (remember that
in the Ashtekar formulation some structure functions appear in the Poisson
brackets of the hamiltonian constraint with itself).

The structure of the Lie algebra is very simple: it is the semidirect sum of
the two Lie algebras $D$ and $W$, where $D$ is the algebra of diffeomorphisms
of
the spatial manifold and $W$ the direct sum of the Lie
algebras of internal gauge rotations and the "dilatations" generated by the
scalar constraint.

At this point it is interesting to note that in 3+1 gravity
\cite{300} there are several scalar densities of weight two that we can
construct with the fields present in the theory and use as the hamiltonian
constraint (given, of course, that the new systems of constraints thus formed
are first class). At variance with the 3+1 case, in which some of the new
constraints describe theories that are not sectors of general relativity,
we will see that the constraint surfaces described by the constraints defined
below in the 2+1 case,
are contained in the hypersurface given by the 2+1 Ashtekar constraints.

We start by noting that in addition to
$\tilde{v}^{I}=\tilde{\eta}_{ab}F_{ab}^{I}$ there is another quantity
with the same index structure and density weight that can be built out of the
triads and connections, namely: $\tilde{w}^{I}=\;\eut_{ab}\epsilon
^{IJK}\tilde{E}^a_J\tilde{E}^b_K$. There are three scalar densities of weight
+2 that can be constructed with $\tilde{v}^{I}$ and $\tilde{w}^{I}$ only:
$\tilde{v}^{I}\tilde{v}_{I}$, $\tilde{v}^{I}\tilde{w}_{I}$, $\tilde{w}^{I}
\tilde{w}_{I}$. The first two give rise, simply,
to the constraints (\ref{2}) and
(\ref{2c}). The third  gives the constraint $\tilde{w}^{I}\tilde{w}_{I}=0$
which is just the  condition that the (double-densitized) 2-metric built
with the triads $\qu^{ab}$ is degenerate (that is $\det \qu=0$). As this last
constraint is a function only of the triads it's
Poisson bracket with itself will
be zero and adding it to the Gauss and vector constraints we will have a first
class system:

\begin{eqnarray}
& & {\cal D}_{a} \tilde{E}^{a}_{I}=0\nonumber\\
& & \tilde{E}^{b}_{I} F_{ab}^{I}=0\label{233}\\
& & \epsilon^{IJK}\epsilon_{I}
^{\;\;LM}\;\eut_{ab}\;\eut_{cd}\tilde{E}^a_J\tilde{E}^b_K\tilde{E}^c_L
\tilde{E}^d_M=0\nonumber
\end{eqnarray}

We now show
that (\ref{233}) describe the degenerate sectors of the Ashtekar constraints
(\ref{2}). We start by noting that there are two possible ways of having
$\tilde{w}^{I}\tilde{w}_{I}=0$: either because $\tilde{w}^{I}=0$
(which implies that the triads are colinear) or because $\tilde{w}^{I}$ is
null (in which case the triads must be contained in a null plane orthogonal to
$\tilde{w}^{I}$). If
$\tilde{w}^{I}=0$ then $\tilde{E}^1_I$ and $\tilde{E}^2_I$ are colinear and
the vector constraint tells us that both must be orthogonal to $\tilde{v}^{I}$.
Now $\tilde{v}^{I}$, $\tilde{E}^1_I$ and $\tilde{E}^2_I$ are contained in a
plane and the hamiltonian constraint in (\ref{2}) is satisfied. If
$\tilde{w}^{I}$ is null and different from zero then the vector constraint
tells us that $\tilde{w}^{I}$ and $\tilde{v}^{I}$ must be proportional to each
other (and
hence both null) so that $\tilde{v}^{I}$, $\tilde{E}^1_I$ and $\tilde{E}^2_I$
are contained in a plane and again the scalar constraint in (\ref{2}) is
satisfied.\footnote{It is worth pointing out now that it is not possible to
define a new "vector constraint" $\tilde{w}^{I}\tilde{E}^a_I=0$ because this
is just an identity.} Introducing:

\begin{equation}
S_D(\nut)=\frac{1}{2}\int_{\Sigma}d^{2}x \nut
\epsilon^{IJK}\epsilon_{I}
^{\;\;LM}\;\eut_{ab}\;\eut_{cd}\tilde{E}^a_J\tilde{E}^b_K\tilde{E}^c_L
\tilde{E}^d_M
\label{333}
\end{equation}

\noindent we see that (\ref{11}) and (\ref{12}) are now substituted by:

\begin{eqnarray}
& & \left\{
S_D(\nut),A^{J}_{a}\right\}=2\nut\epsilon_I^{\;\;JK}\tilde{w}^I\;\eut_{ab}
\tilde{E}^{b}_K
\label{433}\\
& & \left\{S_D(\nut),\tilde{E}_{I}^{a}\right\}=0
\label{533}
\end{eqnarray}

\noindent and the infinitesimal gauge transformations generated by
(\ref{333}) are:

\begin{eqnarray}
& & \left\{ \begin{array}{l}A_{a}^{I}\rightarrow A_{a}^{I}+2\delta\nut
\epsilon_I^{\;\;JK}\tilde{w}^I\;\eut_{ab}\tilde{E}^{b}_K \\
\tilde{E}_{I}^{a}\rightarrow\tilde{E}_{I}^{a}
\end{array}\right.\label{633}
\end{eqnarray}

\noindent Notice that (\ref{533}) implies that the action of the constraints
on the triads reduces to gauge rotations and diffeomorphisms. The degenerate
character of the triad is thus mantained under the evolution given by
(\ref{233}). Finally we point out that the Poisson algebra in this case
coincides with (\ref{133}).

\section{The Degrees of Freedom of 2+1 Gravity}

In this section we will concentrate on the dynamics defined by the Ashtekar
constraints (\ref{2}) and discuss why the existence of degenerate solutions
changes the number of degrees of freedom of the theory. The role of null
curvatures will be essential in what follows. We start by defining
what we will refer to as a "patch". Let R$_{t}$ be a
2-dimensional connected region contained in $\Sigma_{t}$ (which is the
2-dimensional  slice at time 't'). If $\tilde{v}^{I}$ is strictly  non-zero
and null over R$_{t}$ then R$_{t}$ is defined to be a null patch.
Similarly one defines  spacelike, timelike and flat patches.

Consider now initial data which are such that $\Sigma_0$
(the initial surface ) is the union of null and flat patches. Then, if it is
assumed that the evolution of such data is well defined,
we claim that these null  and flat patches cannot be destroyed or created.
In fixed coordinates
on $\Sigma$ these patches may change shape and size and move around $\Sigma$
but
can never appear or disappear.

To prove this claim one must examine the evolution equations.
Evolution is generated by the constraint functionals which we display below:

\begin{eqnarray}
& & G(\Lambda^{I})=\int_{\Sigma}d^{2}x\; \Lambda^{I}
{\cal D}_{a}\tilde{E}^{a}_{I} \\
& & D(N^{a})=\int_{\Sigma}d^{2}x N^{a}[F_{ab}^{I}\tilde{E}_{I}^{b}-
A_{a}^{I}{\cal D}_{b}\tilde{E}^{b}_{I}]\\
& & S_A(\nut)=\frac{1}{2}\int_{\Sigma}d^{2}x \nut
\tilde{E}^{a}_{I}\tilde{E}^{b}_{J}F_{abK}\epsilon^{IJK}
\end{eqnarray}

\noindent The equations of motion are :

\begin{eqnarray}
& & \dot{\tilde{E^{b}_{I}}}=\epsilon_{I}\;^{ LK}\Lambda_{L}\tilde{E}^{b}_{K}
+{\cal L}_{\vec{N}}\tilde{E}^{b}_{I}+{\cal D}_{c}
(\nut\tilde{E}^{c}_{J}\tilde{E}^{b}_{K}\epsilon^{JK}\;_{I}) \\
& & \dot{A_{b}^{I}}=-{\cal D}_{a}\Lambda^{I}
+{\cal L}_{\vec{N}}A_{b}^{I}+\nut\epsilon^{IKL}\tilde{E}^{d}_{K}F_{bdL}
\end{eqnarray}

Given a particular patch, internal SO(2,1) gauge transformations
cannot change the timelike, spacelike, null or flat character of
$\tilde{v}^{K}$, since  $\tilde{v}^{K}$ is just rotated in internal space by
such transformations. Furthermore, smooth diffeomorphisms will only move
patches around on the 2-surface and (in some fixed coordinates) change their
shape and size, but cannot make them appear or disappear.

Hence the only gauge transformations which could possibly alter the number and
type of patches are gauge transformations generated by the scalar constraint.
Let us focus on these:

\begin{eqnarray}
& & \dot{\tilde{E^{b}_{I}}}={\cal D}_{c}
(\nut\tilde{E}^{c}_{J}\tilde{E}^{b}_{K}\epsilon^{JK}\;_{I})\label{400}\\
& &
\dot{A_{b}^{I}}=\nut\epsilon^{IKL}\tilde{E}^{d}_{K}F_{bdL}\label{401}
\end{eqnarray}

Now, consider an initial data set composed of flat and null patches.
In a flat patch (\ref{401}) implies that
$\dot{F_{ab}}=0$.

Let us now consider a null patch;that is,
$\tilde{v}^{I} =\frac{1}{2}F_{ab}\tilde{\eta}^{ab}$ is null and non-zero.
We have now:

\begin{equation}
\dot{A_{b}^{I}}=\nut\epsilon^{IKL}\tilde{E}^{d}_{K}F_{bdL}=
\nut\epsilon^{IKL}\tilde{E}^{d}_{K}\;\eut_{bd}\tilde{v}_{L}
\label{402}
\end{equation}

\noindent In order to proceed we show that in this case it is always possible
to find a vector field $w^{c}$ on $\Sigma$ such that
$v_b^I=\nut\epsilon^{IKL}\tilde{E}^{d}_{K}\;\eut_{bd}\tilde{v}_{L}=
w^{c}\tilde{v}^{I}\;\eut_{cb}$

\noindent We prove first that $v_{b}^{I}\;v_{cI}=0$ by
substituting for $v_{b}^{I}$ from above:
\begin{eqnarray}
v_{b}^{I}\;v_{cI} & = &
\frac{1}{4}\nut^{2}\epsilon^{IKL}\epsilon_{IMN}\tilde{E}^{d}_{K}\tilde{E}^{gM}
   \;\eut_{bd}\;\eut_{cg}\tilde{v}^{N}\tilde{v}_{L} \\
& = & \frac{1}{4}\nut^{2}[ \tilde{E}^{d}_{I}\tilde{E}^{gI}
\tilde{v}^{L}\tilde{v}_{L}-\tilde{E}^{d}_{N}\tilde{v}^{N}
\tilde{E}^{g}_{L}\tilde{v}^{L} ]\;
\eut_{bd}\;\eut_{cg}
\end{eqnarray}
The first term in the square brackets is zero because $\tilde{v}^{I}$is null
and the second term in the square brackets is zero because of the vector
constraint. It is trivial now to see that
$v_{b}^{I}\tilde{v}_{I}=0$
because $v_{b}^{I}\tilde{v}_{I}=
\nut\tilde{E}^{d}_{K}\;\eut_{bd}\epsilon^{IJK}\tilde{v}_{J}\tilde{v}_{I}=0$.
But if $v_{b}^{I}\tilde{v}_{I}$=0 then $v_{b}^{I}$ for b=1,2 lie in the null
(internal) plane. Also, since $v_{b}^{I}\;v_{cI}=0$ these vectors must be
proportional to $\tilde{v}_I$. We conclude then that $v_{b}^{I}=\eut_{cb}w^{c}
\tilde{v}^{I}$ for some vector field $w^{c}$.

\noindent Using the previous result we see that:

\begin{eqnarray}
\dot{A}_{b}^{I} & = & w^{c}\tilde{v}^{I}\eut_{cb} \label{314}\\
                & = & w^{c}F_{cb}^{I}  \label{315}  \\
                & = &
{\cal L}_{\vec{w}}A_{b}^{I}\;-\;{\cal D}_{b}(w^{a}A_{a}^{I})\label{316}
\end{eqnarray}

\noindent Hence the action of the scalar constraint on null patches reduces
to a diffeomorphism plus an internal gauge rotation. Although the vector field
$w^a$ is  defined above only for null patches,  one can easily see that
(\ref{315}) is also true if
$F_{cb}^{I}=0$  for arbitrary $w^a$. Since every null patch is adjacent to
a flat patch, we can choose to extend $w^a$ smoothly from the null patch to the
adjacent flat patch  in such a way that $w^a$ goes to zero smoothly as fast as
we want. In this way we see that we can describe the evolution of flat and null
patch data as diffeomorphisms and gauge rotations generated by
a single smooth vector field $w^a$ via (\ref{316}) (In other words, we have
managed to smoothly extend $w^a$s defined on each null patch by appropriately
extending them into the flat patches so that all  the $w^a$'s so defined can be
smoothly glued together into one smooth vector field on the 2--manifold).
Since  neither  gauge rotations nor
diffeomorphisms can change the null  or flat character of the curvature
we see that these patches can never disappear (or by time reversed
evolution, appear) by evolution generated by the constraints.
Now consider  closed curves completely contained in the flat patches such that
they cannot be deformed to a point without crossing a null patch. We will say
that two such curves are equivalent if they can be continously deformed into
one another without crosing a null patch. There is no
reason, in principle, to expect that the holonomies of the connection
around these curves are trivial. In fact it is true (as we shall see in the
next
section) that in the presence of an arbitrary number of null and flat
patches we can freely specify the
holonomies of the connections around inequivalent curves.

Note that the traced holonomies along the curves in flat patches are
observables, i.e. they commute with the Ashtekar constraints (one can easily
check this by first computing the poisson brackets with the constraints and
then setting $F_{ab}^{I}=0$).
If we can construct initial data corresponding to an arbitrary number of
null and flat
patches (and if this data depends  on a finite number
number of parameters $per\;\;patch$ with a gauge
invariant interpretation, i.e. the holonomies considered above) and if the
evolution of such  initial data is well defined then we can
conclude that there are infinitely many gauge inequivalent solutions to the
theory. The need for an arbitrary number of new parameters
to describe the gauge inequivalent classes of solutions tells us that there
is an arbitrary number of degrees of freedom in the degenerate sectors of 2+1
gravity described by the Ashtekar constraints, i.e. the reduced phase space in
this case is infinite dimensional. In order to identify all the degrees of
freedom of the sector of the theory described by the "patch" solutions
one must also study what happens inside the
non-flat patches and see how many parameters one needs
 in order to completely describe the reduced phase space.

\section{Existence of Patch Initial Data}

In this section we exhibit initial data composed of patches in the case in
which the two dimensional
 slices have the topology of a torus. These solutions to the
constraints depend on an infinite number of real parameters with a gauge
invariant meaning (they are traces of holonomies around curves contained in
the flat patches). We prove then, by giving an explicit example, that the
reduced phase space for the Ashtekar description of 2+1 gravity is infinite
dimensional\footnote{Notice that we are not giving a complete description of
the reduced phase space, but only showing that it is infinite dimensional by
constructing an infinite dimensional subset contained in it}

We begin by fixing coordinates ($\theta, \phi$) on the torus $T=S^1\times S^1$.
with values in the interval $[0,2\pi]$ (we identify points corresponding to 0
and $2\pi$).
We also fix an internal orthonormal basis $(x_{I}, y_{I}, t_{I})$
in the Lie algebra of SO(2,1)  such that $x_{I}, y_{I}$ are spacelike and
$t_{I}$ is timelike and make the following ansatz for
the phase space variables in
our coordinates:
\begin{eqnarray}
\tilde{E}^{\theta}_{I}=E_{1}x_{I}  &    &   A_{\theta}^{I}=A_{1}x^{I}\\
\tilde{E}^{\phi}_{I}=E_{2}y_{I}+E_{3}t_{I} &   &
A_{\phi}^{I}=A_{2}y^{I}+A_{3}t^{I}
\end{eqnarray}
where $E_{1},E_{2},E_{3},A_{1},A_{2},A_{3}$ are functions only of the $\theta$
coordinate. Putting this ansatz into the constraint equations, we find that
the nontrivial constraints we get, are the scalar constraint, the vector
constraint in the $\theta$ direction and the Gauss constraint in the internal
'x$_I$' direction. The rest of the vector and Gauss constraints are trivially
zero. The constraints are:\\

\begin{eqnarray}
& & E_{1}E_{2}(A_{3}'-A_{1}A_{2})- \
E_{1}E_{3}(A_{2}'-A_{1}A_{3})=0\nonumber\\
& & E_{2}(A_{2}'-A_{1}A_{3})- E_{3}(A_{3}'-A_{1}A_{2})=0\label{555} \\
& & E_{1}'+(A_{2}E_{3}-A_{3}E_{2})=0\nonumber
\end{eqnarray}

(In the above set of equations and elsewhere in this section a prime means a
derivative with respect to $\theta$). The only non-zero component of the
curvature is:

\begin{equation}
F_{\theta\phi}^I=(A_{2}'-A_{1}A_{3})y^{I}+(A_{3}'-A_{1}A_{2})t^{I}\label{556}
\end{equation}

\noindent so that it is flat when $A_{2}'-A_{1}A_{3}=A_{3}'-A_{1}A_{2}=0$ and
null when $A_{2}'-A_{1}A_{3}=\pm(A_{3}'-A_{1}A_{2})$. We construct a solution
to the constraints with an arbitrary number N of flat and null patches. We will
start by defining the patches and then constructing a connection
adapted to them (i.e. flat in the flat patches and null in the null ones). We
will then use the constraints and the above connection to build the triads.

The $i^{th}$ null patch extends from $\theta=a_{i}$ to $\theta=b_{i}$. There
are N-1 flat patches in the regions $\theta=b_{i}$  to $\theta=a_{i+1}$ for
$i=$1 to N-1 and one flat patch which covers the region $(b_{N}, 2\pi)
\cup(0, a_{1})$.
Here $a_{i},b_{i}$ are values of $\theta$ such that
$0< a_{1}< b_{1}< a_{2}<...<a_{N}<b_{N}<2\pi$.

In the following we will take $A\equiv A_{1}=A_{2}$. It is straightforward to
check that if we take a smooth $A(\theta)$ such that it is arbitrary in the
null patches and zero in the flat ones then $A_{3}$ must be:

\begin{equation}
A_{3}(\theta)=A(\theta)+c_{0}\exp
\left\{-\int_{0}^{\theta}A(\xi)d\xi\right\}\;\;c_{0}\neq 0\label{557}
\end{equation}

\noindent If $\theta$ is in the flat patch between $b_{i}$ and $a_{i+1}$ it is
clear that $A_{3}$ is a constant $c_{i}$ (we will impose that $c_{i}\neq 0$).
As a consequence of the matching condition $A_{3}(0)=A_{3}(2\pi)=c_{0}$ we
have the following condition on $A(\theta)$:

\begin{equation}
\int_{0}^{2\pi}A(\xi)d\xi=0\label{558}
\end{equation}
\noindent An example with four null and four flat patches is shown in figure 1.
Until now no use of the constraints has been made; we will use them in
order to find the triads. In the flat patches the constraints reduce to the
single equation:

\begin{equation}
E_{1}'+A_{2}E_{3}-A_{3}E_{2}=0\label{559}
\end{equation}

\noindent  whereas in the null ones we have:

\begin{eqnarray}
& & E_{2}=E_{3}\label{560}\\
& & E_{1}'+(A-A_{3})E_{2}=0\label{561}
\end{eqnarray}

\noindent Notice that the metric is given by:

\[ \qu^{ab}=\left[ \begin{array}{cc}
E_{1}^{2} & 0 \\ 0 &
E_{2}^{\;\;2}-E_{3}^{\;\;2}
\label{562}
\end{array} \right] \]

\noindent and then the constraints imply that in the null patches $\qu^{ab}$ is
degenerate. In the $i^{th}$ flat patch we have $A=0$ and $A_{3}=c_{i}$ and then
(\ref{559}) reduces to $E_{1}'=c_{i}E_{2}$. If we choose $E_{2}=1/c_{i}$ then
we have $E_{1}=\theta+p_{i}$ for $i=1,...,N-1$. In the two parts of the flat
patch which covers $(b_{N}, 2\pi)\cup(0, a_{1})$ we will write
$E_{1}=\theta+p_{0}$ and $E_{1}=\theta+p_{N}$ respectively. The matching
condition $E_{1}(0)=E_{1}(2\pi)$ tells us then that $p_{0}=2\pi+p_{N}$. As
there
are no restrictions on $E_{3}$ we will choose it such that $E_{2}\neq E_{3}$,
$E_{2}(a_{i})=E_{3}(a_{i})$ and $E_{2}(b_{i})=E_{3}(b_{i})$; with this choice
the metric is non-degenerate in the flat patches. In order to complete the
construction of the triads we extend $E_{1}$ to the null patches in a smooth
way and get:

\begin{equation}
E_{2}=E_{3}=\frac{E_{1}'}{A_{3}-A}\label{563}
\end{equation}

\noindent Notice that $E_{2}$ and $E_{3}$ are well defined because $A\neq
A_{3}$ (see figure 2).

Consider now the $i^{th}$ flat patch. Let the loop going from $\phi=0$ to
$\phi=2 \pi$  at some fixed value of $\theta$ $in\; the \;flat \;patch$ be
called $\gamma_{i}$. Then
\begin{equation}
T^{0}_{\gamma_{i}}(A):= Trace({\cal P}
\exp(\int_{\gamma_{i}}A_{\phi}d\phi))\label{564}
\end{equation}

\noindent is left invariant under the "evolution" defined by the constraints
($T^{0}_{\gamma_{i}}(A)$ is just the trace of the holonomy of the connection
on the loop $ \gamma_{i}$. The trace is performed with respect to a
2-dimensional representation of SU(1,1))\\
By direct calculation:
\begin{center}
$T^{0}_{\gamma_{i}}(A)=2\cos(\pi c_{i})$
\end{center}

We have thus exhibited a family of solutions to the Ashtekar constraints of 2+1
gravity parametrized by the gauge invariant quantities $T^{0}_{\gamma_{i}}(A)$
that can be continuously changed by varying the $c_{i}$. As the number of
constants $c_{i}$ is arbitrary we conclude that there are regions of
arbitrarily large dimension in the reduced phase space defined by the
Ashtekar constraints of 2+1 gravity.

\section{Conclusions}

Let us first summarize our results. From the observation  that null
curvatures play a very special role in the  Ashtekar description of 2+1
gravity,
we have obtained two interesting sets of results. The first one
is the finding of two new systems of constraints that describe some sectors of
the  constraint hypersurface of the Ashtekar theory. One of these systems of
constraints is a generalization of the Witten constraints of the theory whereas
the other (i.e. Ashtekar constraints) describes its  degenerate sectors.
The second result of the paper is
the realization that the Ashtekar constraints actually describe a system with
an infinite number of degrees of freedom.

We now make some comments about these results:

\noindent (a) It should be mentioned that though we have shown the reduced
phase
space is infinite dimensional, we do not know where our solutions lie in the
full reduced phase space of the theory. In particular they may be disconnected
from the Witten sector of the reduced phase space or might be in some
pathological non manifold part of the reduced phase space.

\noindent (b) As we have seen, inclusion  of degenerate metrics  in the
Ashtekar
description of 2+1 gravity leads to surprising results. Hence one should think
carefully about whether one wants to include degenerate metrics in the
classical description  of 3+1 gravity. Here issues get complicated by the fact
that nontrivial reality conditions are to be imposed \cite{jose}.

\noindent (c) It would be of great interest to try and quantize the 2+1
Ashtekar
theory. The theory from the analysis in this paper seems to be very
complicated.As a first step Manojlovi\'c and Mikovi\'c \cite{nenad}
have looked at a homogeneous
minisuperspace of the theory on the torus (It is our understanding that
they claim to gauge fix the full theory. But this cannot be so because we know
that the theory has an infinite number of degrees of freedom). It turns out
that within this homogeneous minisuperspace the theory  has different sectors.
Manojlovi\'c and Mikovi\'c have quantized only the timelike curvature sector in
their analysis. We are currently investigating the other sectors of the
homogeneous minisuperspace.

\noindent (d) It may seem that inclusion of degenerate metrics unnecessarily
complicates the Ashtekar description of of 2+1 gravity. One could  take the
viewpoint that as far as classical physics goes, degenerate metrics are
irrelevant (note that when the metric is nondegenerate the Ashtekar description
is equivalent to the geometrodynamic description in 2+1 as well as 3+1
dimensions). But then the question naturally arises as to how to proceed with
the quantization. Are we simply back to a quantization in terms of
geometrodynamic variables? One interesting way of answering  "no" to the above
question is to construct a loop representation for the Ashtekar description of
2+1 gravity just as in 3+1 gravity \cite{lee}. This is still an open problem.

\noindent (e) Within the context of vaccuum Ashtekar
2+1 gravity itself, it turns out that in the
case of a non-compact 2-manifold, namely the 2-dimensional plane, it is
possible
to construct solutions similar to the "patch" data presented in this paper,
on which the evaluated mass can be anything we want i.e. positive or negative
(note that the mass is just proportional to the trace of the holonomy along
a loop
going along the $S^1$ at spatial infinity). This raises the issue of whether
these solutions can simulate particles.

\noindent (f) Although we have described patch solutions involving only null
and
flat patches, it is possible to construct initial data involving other types of
patches too. But we have not analyzed what happens to such initial data under
evolution.

In conclusion,  we note that contrary to naive expectation the 2+1 Ashtekar
description has a very nontrivial structure and can serve as a useful toy model
for the Ashtekar quantization program in 3+1 dimensions.

\vspace{1cm}

{\bf Acknowledgements}: The authors want to thank A. Ashtekar, A. Corichi,
V. Husain, G. Immirzi, J. Louko, N Manojlovi\'c,  G. Mena, H. Morales,
T. Schilling, L. Smolin, and R. Sorkin for several interesting discussions and
remarks. J.F.B.G. is supported by a C.S.I.C. postdoctoral fellowship.

\newpage

\noindent{\bf Figure Captions}

\vspace{2cm}

\noindent {\bf Figure 1} Plots of $A(\theta)$ (a) and $A_{3}(\theta)$ (b) with
four null patches $\;(\pi/4, \pi/2)$, $(3\pi/4, \pi)$, $(5\pi/4, 3\pi/2)$,
$(7\pi/4, 2\pi)\;$ , and four flat patches.

\noindent {\bf Figure 2} Plots of $E_{1}(\theta)$ (a), $E_{2}(\theta)$ (b) and
$E_{3}(\theta)$ (c). These triads describe a metric $\qu^{ab}$ that is
non-degenerate in the flat patches and has signature (+, +).


\newpage

\begin{thebibliography}{33}

\bibitem{1} A. Ashtekar, {\sl Phys. Rev. Lett. 57 (1986) 2244}

\bibitem{2} A. Ashtekar, {\sl Phys. Rev. D36 (1987) 1587}

\bibitem{100} S.Deser, R. Jackiw and G. 't Hooft, {\sl Ann. Phys. (NY) 152
(1984)220}

\bibitem{4} E. Witten, {\sl Nucl. Phys. B311 (1988) 46}

\bibitem{3} A. Ashtekar, V. Husain, C. Rovelli, J. Samuel and L. Smolin, {\sl
Class. Quantum Grav. 6 (1989) 185};

A. Ashtekar and J. Romano, {\sl Phys. Lett. 229B (1989) 56}

\bibitem{5} I. Bengtsson, {\sl Phys. Lett. 220B (1989) 51}

\bibitem{6} M. Varadarajan, {\sl Class. Quantum Grav. 8 (1991) L235}

\bibitem{300} I. Bengtsson, {\sl Phys. Lett. 254B (1991) 55}

\bibitem{jose} J.W. Maluf, {\sl Class. Quantum Grav. 10 (1993) 805};

G. Immirzi, {\sl to appear in Class. Quantum Grav.}

\bibitem{nenad} N. Manojlovi\'c and A. Mikovi\'c, {\sl Nucl. Phys. B 385 (1993)
571}

\bibitem{lee}   C. Rovelli and L. Smolin, {\sl Nucl. Phys. B 331 (1990) 80}

\end{thebibliography}
\end{document}